# Self-deflecting plasmonic lattice solitons and surface modes in chirped plasmonic arrays

Chunyan Li,[1] Ran Cui,[2] Fangwei Ye,[1,*] Yaroslav V. Kartashov,[3,4] Lluis Torner,[3] and Xianfeng Chen[1]

[1]State Key Laboratory of Advanced Optical Communication Systems and Networks, Department of Physics and Astronomy, Shanghai Jiao Tong University, Shanghai 200240, China
[2]School of Biomedical Engineering, Shanghai Jiao Tong University, Shanghai 200240, China
[3]ICFO-Institut de Ciencies Fotoniques, and Universitat Politecnica de Catalunya, Mediterranean Technology Park, 08860 Castelldefels (Barcelona), Spain
[4]Institute of Spectroscopy, Russian Academy of Sciences, Troitsk, Moscow Region 142190, Russia
*Corresponding author: fangweiye@sjtu.edu.cn

We show that chirped metal-dielectric waveguide arrays with focusing cubic nonlinearity can support plasmonic lattice solitons that undergo self-deflection in the transverse plane. Such lattice solitons are deeply-subwavelength self-sustained excitations, although they cover several periods of the array. Upon propagation, the excitations accelerate in the transverse plane and follow trajectories curved in the direction in which the separation between neighboring metallic layers decreases, a phenomenon that yields considerable deflection angles. The deflection angle can be controlled by varying the array chirp. We also reveal the existence of surface modes at the boundary of truncated plasmonic chirped array that form even in the absence of nonlinearity.

One of the central goals of current nanooptics is the elucidation of strategies that enable engineering and control of the propagation of light in strongly localized waveguiding structures. Frequently, such strategies rely on periodic or aperiodic waveguide arrays, since the coupling rate between adjacent waveguides in such structures, which determines the rate at which light expands across the arrays, can be engineered (for recent reviews, see Refs. [1-3] and references therein). The majority of earlier works on light propagation in periodic media addressed dielectric structures, but there is a growing current interest in metal-dielectric waveguide arrays. Such interest is motivated by the properties of the surface plasmon excitations supported by metal-dielectric materials, which afford the light concentration and guidance at subwavelength scales envisaged for miniaturized devices [4-16]. For example, plasmonic waveguide arrays have been used to focus incident wide waves into a single slit [4], to observe plasmonic Bloch oscillations [4-6] and Zener tunneling [8]. More recently, plasmonic routing was reported in aperiodic graphene arrays [9]. If the plasmonic periodic nanostructure exhibits a nonlinear optical response, the formation of a rich family of self-sustained subwavelength excitations becomes possible [17-28]. Nevertheless, such excitations are usually strongly pinned to the particular dielectric layer that exhibits a nonlinear response, while many potential applications need controllable routing of light across the array.

In this Letter we address the propagation of subwavelength light beams in one-dimensional arrays of metal-dielectric layers, where the separation between the adjacent metallic layers changes linearly across the array. The linear chirp may lead to large transverse drifts of the light beams propagating inside the structures. In addition, if the dielectric host medium exhibits a focusing cubic (Kerr) nonlinearity, one can achieve formation of self-deflecting plasmonic lattice solitons that maintain their width and structure upon propagation even in the case of considerable self-deflection angles. We also consider truncated chirped arrays, to show that the boundaries of such arrays act as attractors for light and thus allow formation of surface modes even in the absence of nonlinearity.

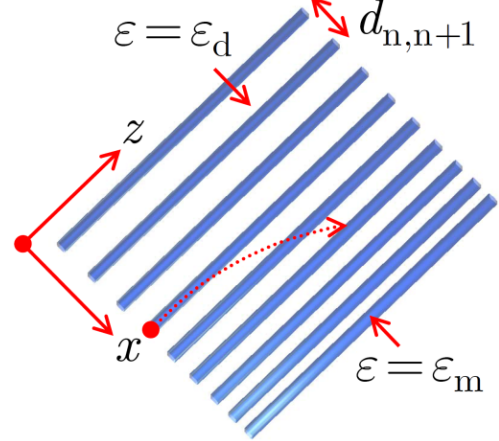

Fig. 1. Schematic representation of a chirped subwavelength array consisting of dielectric layers (white regions) separated by metallic layers (blue regions) of width $a$. The separation between centers of metallic layers $n$ and $n+1$ decreases with layer index $n$ linearly as $d_{n,n+1} = d_{1,2} - (n-1)\delta d$, where $d_{1,2} = 40$ nm. The dotted arrow indicates the propagation and deflection direction of light in the structure.

The chirped metal-dielectric array is sketched in Fig. 1. The structure contains many periods, hence initially no boundary effects are taken into account in the transverse $x$ direction. The structure involves closely spaced parallel metallic (silver) nano-layers with identical widths $a = 20$ nm and a relative dielectric permittivity $\varepsilon_m = -20 - 0.19i$ at the wavelength $\lambda = 632$ nm [29]. A chirp is introduced into the structure by assuming that the separation between layers $n$ and $n+1$ decreases linearly with a constant rate $\delta d > 0$, namely, $d_{n,n+1} = d_{1,2} - \delta d(n-1)$, where $n = 1, 2...$ and $d_{1,2} = 40$ nm. Such a chirp induces asymmetric coupling between layers of the structure and stimulates transverse beam displacements. Metallic layers are embedded into nonlinear dielectric host medium with relative permittivity $\varepsilon_d = (n_0 + n_2 I)^2$, where $n_0 = 3.5$ and $n_2 = 4 \times 10^{-18}$ m$^2$/W are linear and nonlinear refractive indices, and $I$ is the light intensity. Here we do not take into account nonlinearity of the metal, assuming it is small.

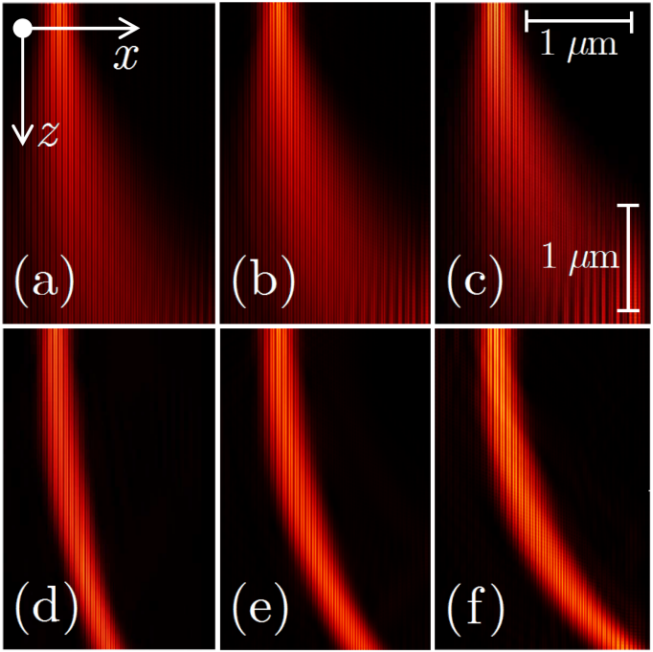

Fig. 2. Evolution dynamics in chirped subwavelength waveguide arrays in the linear (a)-(c) and nonlinear (d)-(f) regimes, in the absence of metallic losses. The chirp rate is $\delta d = 0.5$ nm (a),(d), $0.6$ nm (b),(e), and $0.7$ nm in (c),(f). The propagation distance is $3$ $\mu$m, while the width of the transverse window is $2$ $\mu$m.

The propagation of a TM-polarized (i.e. only $E_x, E_z, H_y$ components of the electric and magnetic fields are nonzero) light beam along the $z$-axis in the chirped plasmonic array is governed by the reduced system of Maxwell's equations,

$$i\frac{\partial E_x}{\partial z} = -\frac{1}{\varepsilon_0 \omega}\frac{\partial}{\partial x}\left(\frac{1}{\varepsilon}\frac{\partial H_y}{\partial x}\right) - \mu_0 \omega H_y, \quad i\frac{\partial H_y}{\partial z} = -\varepsilon_0 \varepsilon \omega E_x, \quad (1)$$

where $\varepsilon_0$ and $\mu_0$ are the vacuum permittivity and permeability, $\omega$ is the light frequency, $\varepsilon(x)$ is the relative permittivity of the chirped array. The system of Eqs. (1) was solved with a finite-element method allowing inclusion of nonlinear effects [30]. Throughout this Letter the input conditions for Eqs. (1) were selected in the form of superposition of linear eigenmodes of the individual metallic layers, surrounded by a dielectric medium, with a Gaussian envelope:

$$E_x(x)|_{z=0} = E_0 \sum_n (-1)^n e_{x,n}(x) e^{-(n-n_c)^2/T^2} \quad (2)$$

where the functions $e_{x,n}(x)$ describe the eigenmodes of the individual metallic layers, $n_c$ determines the number of the layer where excitation is centered, $T$ is the width of the envelope, and $E_0$ is the input amplitude. Metallic layer supports two types of SPP modes: one with symmetric $E_x(x)$ field distribution and the other one having antisymmetric $E_x(x)$ distribution [31]. We use only antisymmetric modes in (2), since they can be found even for very small widths of the metallic layers. The factor $(-1)^n$ guarantees that the excitation has staggered structure in the neighboring layers. Such staggered phase structure is necessary, since we will use focusing nonlinearity to balance beam diffraction [19,21,22]. We set $T = 2.5$ and $n_c = 10$, so that initial excitation covers around 5 layers (its width is approximately 140 nm for $d_{1,2} = 40$ nm and $\delta d = 0.6$ nm) and is centered at the 10th layer [see example in Fig. 5(d)]. Our main goal is to show that for selected parameters of the structure such beams can undergo considerable self-deflection in the transverse plane at the distances $\sim 3$ $\mu$m, while maintaining their width and internal structure and exhibiting minimal attenuation.

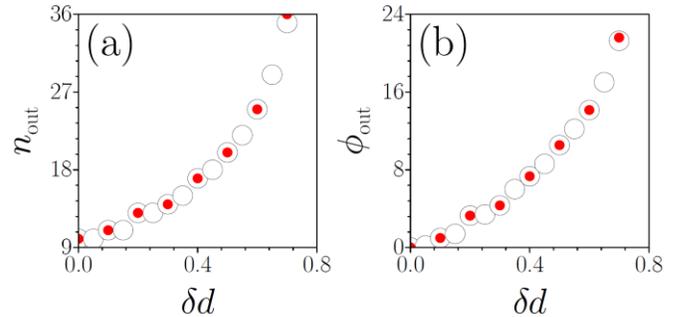

Fig. 3. Output waveguide number (a) and output propagation angle in degrees (b) versus chirp rate $\delta d$ (in nanometers). Empty circles show dependencies obtained in the absence of metallic losses, while red dots correspond to the case with metallic losses. Total propagation distance is $3$ $\mu$m.

Illustrative examples of the propagation dynamics are shown in Fig. 2. To show the effect of self-deflection induced by the lattice chirp, we initially ignored losses in the metallic layers. Since the beam is deeply subwavelength (width $\approx 140$ nm), it undergoes considerable diffraction in the linear case ($E_0 \to 0$) even for a very short propagation distance [$3$ $\mu$m in Figs. 2(a)-2(c)]. The diffraction pattern is strongly asymmetric due to the lattice chirp and the center of mass of the beam shifts in the direction of decreasing separation between metallic layers. The effect of the transverse self-deflection and considerable bending of the propagation trajectory are most apparent in Figs. 2(d)-2(f), where the input peak amplitude of the electric field $E_0 = 5.2 \times 10^9$ V/m was selected such, that the maximal nonlinear contribution to the refractive index in the dielectric region amounts to $\delta n = 0.05$. Strong diffraction of the beam is now nearly completely arrested by the focusing nonlinearity, leading to the formation of subwavelength plasmonic lattice solitons that move across the array, conserving their internal structure and exhibiting acceleration in the transverse plane. Such solitons still cover several me-tallic layers and their widths change only slightly upon propagation. Notice that, in complete contrast to previous works dealing with solitons in plasmonic waveguide arrays [17-28] and their transverse mobility [19], in our case no initial phase tilt is required to set soliton in motion across the array. In other words, the propagation trajectory of such states can be controlled by adjusting the chirp rate of the plasmonic structure.

The dependencies of the position of the output waveguide $n_{\text{out}}$, where the soliton center is located at $z_{\text{out}} = 3\ \mu\text{m}$, and output propagation angle $\phi_{\text{out}}$ on the lattice chirp rate $\delta d$ are shown in Fig. 3. The output deflection angle is approximately defined here as $\phi_{\text{out}} = \arctan(\delta x / z_{\text{out}})$, where $\delta x$ is the transverse beam center displacement acquired upon propagation. Notice that the actual propagation angle defined by the inclination of the tangential line to soliton center trajectory at $z = z_{\text{out}}$ is even higher than $\phi_{\text{out}}$. As Fig. 3 shows, both $n_{\text{out}}(\delta d), \phi_{\text{out}}(\delta d)$ are monotonically growing functions, indicating that larger chirps lead to stronger beam acceleration and deflection. It should be emphasized that our results are obtained using the direct numerical solution of Maxwell's equations, which thus allows large bending angles, in contrast to the paraxial approximation operating with small deflection angles [32].

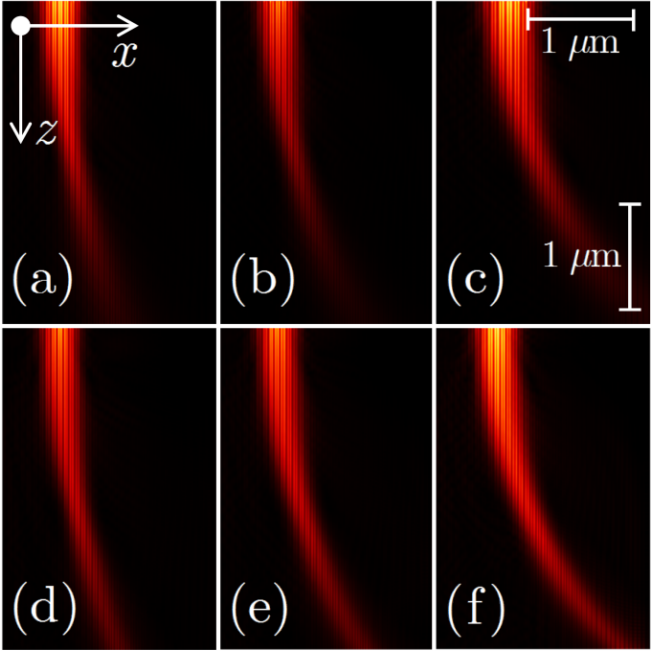

Fig. 4. (a)-(c) Same as in Figs. 2(d)-2(f), but for nonzero losses $\alpha_{\text{m}} = -0.19$ in the metal region. In (d)-(f) gain $\alpha_{\text{d}} = 0.04$ is added into dielectric layers. The propagation distance is $3\ \mu\text{m}$, while the width of the transverse window is $2\ \mu\text{m}$.

A propagation distance of several microns is required in order to achieve large deflection angles. At such distances metallic losses may lead to notable decrease of beam amplitude and also to diffraction [Figs. 4(a)-4(c)]. Nevertheless, we verified that the deflection angle is only weakly affected by metallic losses, as shown in Fig. 3. Moreover, in order to compensate metallic losses one may use active dielectric materials doped with suitable ions. For example, if $\varepsilon_{\text{d}} = (n_0 + n_2 I)^2 + i\alpha_{\text{d}}$, a relatively small gain of $\alpha_{\text{d}} = 0.04$ would be sufficient to nearly compensate losses in silver layers ($\alpha_{\text{m}} = -0.19$ at $\lambda = 632\ \text{nm}$). The corresponding results are shown in Figs. 4(d)-4(f). Note that in the structure considered here, a large fraction of the energy carried by the light beam is concentrated within the dielectric regions and thus the corresponding gain has a stronger impact on the beam evolution than losses inside the metallic layers [25].

Next we address arrays truncated in the direction where separation between metallic layers decreases. We aim at showing that the surface of such an array acts as an attractor of light while total internal reflection at the interface with dielectric material in the point of truncation may result in the formation of stationary surface waves. Thus, we first found stationary surface modes of the linear version of Eqs. (1) in the form $[E_x(x,z), H_y(x,z)] = [E(x), H(x)]e^{ibz}$, where $b$ is the propagation constant determined by the parameters of the structure. The profiles of such waves are depicted in Fig. 5(a) for two chirp rates. Now we fix separation $d_{1,2} = 10\ \text{nm}$ between near-surface layers and let it grow linearly into the depth of the array. The intensity of the surface wave is maximal within the near-surface metallic layer. The localization degree of surface waves strongly depends on the chirp rate and their integral width rapidly decreases with $\delta d$ [Fig. 5(b)]. Such surface waves exist due to the total internal reflection at the edge of the chirped structures, therefore their fundamental origin differs from that of resonant Tamm states, which exist due to Bragg reflection in the periodic structure [28].

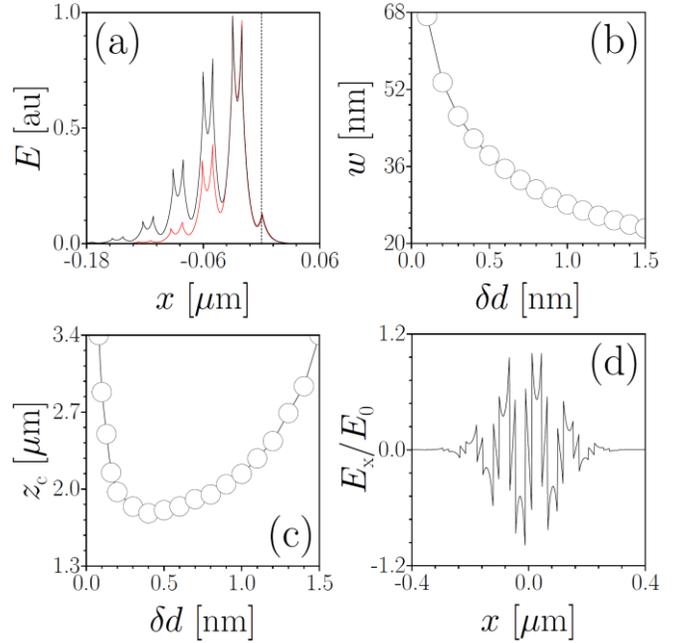

Fig. 5. (a) Profiles of stationary surface modes for $\delta d = 0.4\ \text{nm}$ (black curve) and $\delta d = 1.2\ \text{nm}$ (red curve). (b) Surface wave width versus $\delta d$. (c) Distance of the first collision with the interface for a beam initially centered at the 25th metallic layer, versus $\delta d$. (d) Input used for the excitation of self-deflecting solitons in the bulk of array at $\delta d = 0.6\ \text{nm}$.

Importantly, we verified that the surface waves may be excited by input beams displaced from the interface. This is illustrated in Figs. 6(a),(b), where we use the initial conditions (2) with small $E_0$ and launch light into the second and fifth waveguides (for convenience we now enumerate waveguides from the interface) for a fixed chirp $\delta d = 1.2$ nm. The excitation is most effective when the beam enters the second waveguide, while for larger displacements complex interference patterns occur due to the interference between the light bending toward the interface and the reflected waves. Deflection of the input beam toward the interface is most pronounced when the beam is launched far from the interface. As shown in Figs. 6(d),(e), in such case, a sequence of collisions of subwavelength beams with the interface mediated by completion of bending toward the interface and reflection at the interface takes place. Notice that collisions with the interface can be observed even in the presence of metallic losses [Figs. 6(c),(f)]. The distance $z_c$ of the first collision with the interface is a non-monotonic function of the chirp $\delta d$ [Fig. 5(c)]. Indeed, when $\delta d \to 0$ the array becomes periodic and no deflection toward the interface can occur, while for large $\delta d$ values the metallic slabs become nearly decoupled and do not lead to beam deflection. Collision length becomes minimal for $\delta d \sim 0.4$ nm.

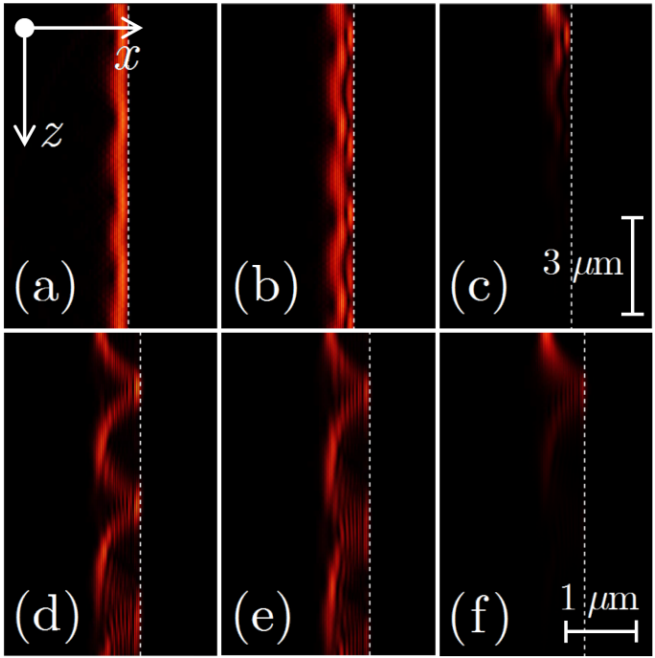

Fig. 6. Evolution of a beam launched into the second (a) and fifth (b),(c) waveguides in a truncated chirped subwavelength array with $\delta d = 1.2$ nm. Evolution of a beam launched into the 25th waveguide for $\delta d = 0.4$ nm (d) and 0.8 nm (e),(f). In (c),(f) metallic losses are taken into account. The propagation distance is 10 $\mu$m and the width of the transverse window is 3 $\mu$m. Dashed lines: position of the interface.

Summarizing, we showed that light beams propagating in suitable chirped plasmonic waveguide arrays may undergo significant transverse deflections. When the host material of the array exhibits a self-focusing nonlinearity, localized plasmonic solitons that accelerate in the transverse plane may form, thus leaving the array at controllable angles that increase with the chirp rate. We also showed that truncated chirped plasmonic arrays may support linear surface waves whose localization depends on the chirp rate.

The work of C. Li and F. Ye is supported by the National Natural Science Foundation of China, Grant No. 11104181 and 61475101.